\documentstyle[aps,eqsecnum,preprint,tighten]{revtex}
\def\lapp{\hbox{$ {     \lower.40ex\hbox{$<$}
 \atop \raise.20ex\hbox{$\sim$}
 }     $}  }
\begin{document}
\draft


\preprint{\vbox{Submitted to Phys.\ Rev.\ C\hfill IU/NTC 93-28\\
                               \null\hfill  FSU-SCRI-93-161\\
                                \null\hfill nucl-th/9312022}}

\title{DYNAMICAL COLOR CORRELATIONS IN A $SU(2)_c$ QUARK EXCHANGE
MODEL OF NUCLEAR MATTER}
\author{S. Gardner\thanks{AAUW Educational Foundation American
Postdoctoral Fellow}$^,$\cite{svg} and C.~J. Horowitz\cite{cjh}}
\address{Nuclear Theory Center and Department of Physics\\
Indiana University, Bloomington, IN 47405, USA}
\author{J. Piekarewicz\cite{jp}}
\address{Supercomputer Computations Research Institute\\
Florida State University, Tallahassee, FL 32306, USA}
\date{December 24, 1993}
\maketitle
\begin{abstract}
The quark exchange model is a simple realization of an adiabatic
approximation to the strong-coupling limit of Quantum Chromodynamics (QCD):
the quarks always
coalesce into the lowest energy set of flux tubes. Nuclear matter is
thus modeled in terms of its quarks. We wish to study the
correlations imposed by total wavefunction antisymmetry when
color degrees of freedom are included. To begin with, we have considered
one-dimensional matter with a $SU(2)$ color internal degree of freedom only.
We proceed by constructing a totally antisymmetric, color singlet
{\it Ansatz} characterized by a variational parameter $\lambda$
(which describes the
length scale over which two quarks in the system are clustered into
hadrons) and by performing a variational Monte Carlo calculation
of the energy
to optimize $\lambda$ for a fixed density. We calculate the $q-q$ correlation
function as well, and discuss the qualitative differences between
the system at low and high density.
\end{abstract}
\pacs{}


\section{INTRODUCTION}
\label{intro}

With the advent of new, high-energy facilities for nuclear physics
research (CEBAF, RHIC), the issue of identifying departures from a
conventional, meson-baryon description of nuclear phenomena
presses.
One would naively expect that as the typical distance scales
probed become short -- relative to the radius of the proton
-- that
the fundamental ingredients of the strong interaction, namely, the
quarks and gluons, should
become manifest in nuclear observables.
This has not been the case.
Indeed, unambiguous signatures of the formation of a
``quark-gluon plasma'' (QGP), hoped for at RHIC, have proven elusive.
The fickleness of many
newly-proclaimed signatures stem from the disparate
pictures of observables in the quark/gluon or
hadronic viewpoints. For example, $J/\psi$ suppression was
heralded as a clear signature of
QGP formation \cite{matsuisatz}, yet it seems that suppression
can also be realized in a hadronic description of high density
matter at high temperatures without plasma formation \cite{gavin,kurihara}.
The divorce of the two pictures is perhaps forced by the difficulty
of treating quark confinement. We offer no new insight into this technical
problem, but, rather, argue that it is useful to consider a simple
model which interpolates between a hadron-based and
quark-based description at low and high density.

In this paper, we shall consider nuclear matter from the viewpoint of
a constituent quark model. The particular model we use, the quark
exchange model, has ``natural'' low and high-density limits:
it behaves as a system of isolated hadrons at low density and
as a non-relativistic
Fermi gas of quarks at high density.
Our purpose is to attempt to understand the
structure of nuclear matter in terms of its quarks. Potentially, we would like
to understand traditional issues, such as nuclear binding and saturation,
from a quark model viewpoint. We anticipate that some ``duality'' may
exist between the quark-gluon and hadronic descriptions. Indeed, the
study of this quark-based description at nuclear matter density should
help us to identify to what extent the descriptions are equivalent.
However, this is not our sole purpose.
In our, albeit simple, model, the structure of the system at fixed density
is controlled entirely
by the quark exchange dynamics, rather than by approximations to the
hadronic and quark-based phases which are neither obviously consistent
nor compatible. Consequently, we regard the study of the structure of
the system with density as the major focus of our study. We hope
to gain insight into the nature of a quark-hadron phase transition, as
well as into the system's structure above and below the
supposed transition point.
In principle, our study also allows us to search for new collective
excitations of the system, which could be novel in structure and unique
to the quark-based description. However,
we shall not explore such venues here.

  The model we use is the quark exchange model \cite{lenz}.
The quark exchange model is a non-relativistic, potential model, with
many-body, confining forces. The many-body nature of the interactions
allows the saturation of the hadron-hadron forces -- there are no
long-range hadron-hadron forces in nature -- to be imposed by fiat.
The original quark exchange model \cite{lenz} was constructed for
the $q^2 \overline{q}^2$ system in three spatial dimensions and
in $U(1)$ color. In this model, the $U(1)$ limit is still confining, and
the system studied shows rich behavior. For example,
the symmetric spatial state
possesses a weakly bound state. The binding energy's weakness,
compared to the hadron confinement scale, is reminiscent of the deuteron.
As a result, this model has been used in several previous quark model
studies of nuclear matter and of finite many-quark systems
\cite{hmn,cjhjp1d,cjhjp3d,watson,wanda,frichter,gardner,lissia}.
For the most part, the studies are confined to one-spatial dimension and
retain the restriction to $U(1)$ color as well. These studies are of
``spin zero fermions'': spin is not explicitly treated, so that the spatial
wavefunctions are required to be manifestly antisymmetric. This has the
unfortunate effect of generating nuclear matter based on hadron clusters in
parity-odd states.
Horowitz and Piekarewicz
have extended their work in one dimension \cite{cjhjp1d}
to three dimensions and have, in addition,
 considered an approximate extension to $SU(3)$ color \cite{cjhjp3d}.
The studies, for the most part, rely heavily
on variational Monte Carlo methods; however,
the authors of Ref.~\cite{wanda}  use a two-body reduction of the many-body
Hamiltonian of Ref.~\cite{lenz}, so that they can use conventional
many-body techniques.

  This work is concerned with the complete treatment of color degrees
of freedom, and with the resulting structure of the system's ground state.
In our model, the gluon degrees of freedom are passive -- they act
only to confine the quarks, so that
they can not be excited. Only the quarks, then, carry an explicit
color label; our treatment is ``complete'' in this context.
As in previous papers, we shall continue to ignore the quarks'
spin. As we do include color, though,
nuclear matter
is based on parity-even hadron clusters, as desired.
A full treatment of color necessarily implies enumeration of all
the exchange terms
required for the construction of a fully antisymmetric state
for a finite number of quarks.
This necessity limits this initial study to $SU(2)$ color
and one spatial dimension, for tractability. We believe, however,
that the techniques presented here can be extended to $SU(3)$ color
and three spatial dimensions.

The simple model we use has many short-comings: it is not relativistic,
even at high density, and there is no mechanism for the production
and propagation of virtual $q\overline q$ pairs. Moreover, chiral
symmetry has no meaning.
Yet, we believe that the model study
presented here gives useful insight into the nature of hadronic
matter as a function of density. The model does include the
effects of confinement and antisymmetry, and its virtue
is that the structure of the system as a function of density is determined
by this physics. To some extent, our results can be understood in
terms of the interplay of these two effects. The qualitative insight
we gain as a result should be relevant in understanding the
predictions of richer models, as the effects we have included,
confinement and antisymmetry, surely play no minor role.

Section~\ref{theqexch} summarizes the $U(1)$ quark exchange model, and
Sec.~\ref{themagmod} describes its extension to $SU(2)$ color.
We compare our model with other models of color in
Sec.~\ref{conferre}, before proceeding, in Sec.~\ref{monte}, to describe
the Monte Carlo computations. Our results are summarized in Sec.~\ref{results},
and we conclude with an outlook towards future work in Sec.~\ref{sumup}.


\section{THE QUARK EXCHANGE MODEL}
\label{theqexch}

  Here we describe the quark exchange model. For two isolated
quarks in one spatial dimension, we consider the Hamiltonian
\begin{equation}
H=-{1\over 2}\partial_1^2 - {1\over 2}\partial_2^2 + {1\over 2}(x_1-x_2)^2 \;.
\label{twoquark}
\end{equation}
The units have been chosen such that the quark mass $m_q$ and the
oscillator frequency $\omega$ are $m_q=\hbar=\omega=1$.
We shall
consider harmonic confinement exclusively, merely for tractability.
The potential term can be thought of as a flux tube which confines quarks
1 and 2. For $N$ quarks, we wish to generalize the potential term so that
confining potentials
``connect'' the quarks to yield the lowest possible potential energy.
We shall avoid
any explicit discussion of color for the moment, and
simply require that the quark ``links'' act in a pair-wise fashion.
Following Horowitz {\it et al.} \cite{hmn},
the Hamiltonian for $N$ quarks becomes
\begin{equation}
H=-{1\over 2} \sum_{i=1}^N \partial_i^2 + V(x_1 \dots x_N) \;,
\label{hamdef}
\end{equation}
where the potential $V(x_1 \dots x_N)$ is
\begin{equation}
 V(x_1 \dots x_N)= \min_{[P]} V_P(x_1 \dots x_N)
\label{minpot}
\end{equation}
such that
\begin{equation}
V_P(x) \equiv \sum_{i=1}^{N/2} v(x_{P^{2i-1}} - x_{P^{2i}})
\end{equation}
with $v(x_{P^1}-x_{P^2})=(x_{P^1}-x_{P^2})^2 /2$.
The $P$ are elements of the permutation group $S_N$, where
$P^i$ is the $i^{\rm th}$ member of the element
$P$. In this way, all possible pair-wise confining potentials are
sampled.

\vbox to 4.5in{\vss\hbox to 5.625in{\hss
{\includegraphics{tahiti.ps}}\hss}}
\nobreak
{\noindent\narrower{{\bf FIG.~\protect{\ref{figtahiti}}}.
An illustration of quark exchange dynamics for 4 quarks.
Two incoming hadrons with quark content $(12)_0$
and $(34)_0$ may exchange quarks -- the
strings connecting 1 to 2  and 3 to 4 flip --
to yield a $(13)_0(24)_0$
outgoing state. }}
\bigskip

As the potential requires that the quarks be paired in a way that the
global potential energy is minimized, the model is driven by quark
exchange dynamics. That is, slight changes in the
spatial configurations of the quarks may result in rearrangements of the
links which confine them. The exchange of quarks between links is the
only dynamical feature of the model. These dynamics are illustrated
schematically in
Fig.~\ref{figtahiti}.

The advantage of the simple prescription of Eq.~(\ref{minpot}) is that
cluster separability is guaranteed. By ``cluster separability,'' we
mean that no long-range, residual forces exist between color-singlet
clusters at large separations. Such color Van der Waals
forces do not exist in nature \cite{feinberg}, so that
cluster separability is a desirable property.
Yet in all models with pair-wise, confining forces, color Van der Waals
forces (dipole-dipole interactions) do exist \cite{lipkin}.
The potential of Eq.~(\ref{minpot}) succeeds in providing cluster
separability because
the forces are of an intrinsically many-body nature.

In some sense, Eq.~(\ref{minpot})
represents an adiabatic approximation to a strong coupling picture of
quark-quark interactions. The gluon degrees of freedom, which are inert,
act as links which connect the quarks to yield the lowest potential energy.
We do not include the possibility of gluonic excitation in our model;
the absence of low-lying
$q\overline q$ states with exotic quantum numbers \cite{paton}
supports this model choice.
The interaction is adiabatic, as the links rearrange themselves
instantaneously in adjustment to the configuration of the quarks.
This picture relies on a separation of time-scale between the quark and gluon
degrees of freedom in QCD. This is not unreasonable for heavy quarks; this is,
fortunately, the same limit in which the non-relativistic quark model,
which is also used here, has validity. For realistic, two-flavor
quark matter, these approximations are not obviously robust; we adopt
them out of necessity.

The non-relativistic quark model has been justified {\it a posteriori}
by its phenomenological success in reproducing a broad
range of baryon properties \cite{isgur}.
In addition, the $U(1)$ $q^2 \overline{q}^2$ model of Lenz {\it et al.}
\cite{lenz} evinces appealing qualitative behavior. As mentioned earlier,
a weak bound state -- compared to the confinement scale -- exists in
three dimensions. This scale is reminiscent of the deuteron.
There also exist hidden channel resonances in the elastic phase shift, which
occur at the hadronic inelastic channel thresholds.
Yet, significant corrections to the adiabatic picture on which
the Lenz {\it et al.} model is based likely exist.
Existing studies focus on
the $N-N$ system, which suggest that the corrections are not small
\cite{harvey,maltman}.
The simple Lenz {\it et al.}
model does have interesting physics content, so that we shall proceed to
explore it in an extension to quark matter in $SU(2)$ color.


\section{MAGIC MODEL IN SU(2) COLOR}
\label{themagmod}

Here we shall consider $SU(2)$ color quark matter in one spatial dimension,
although we shall continue to ignore the quarks' spins.
We want to extend the model above in such a way that the ground state
is guaranteed to be an overall color singlet and that
the system reduces at low density to color-singlet hadron
clusters. These constraints are not sufficient to specify the
extension to $SU(2)$ color uniquely. Starting with the $U(1)$ potential
of Eq.~(\ref{minpot}), one possible extension is
\begin{equation}
 V_{\rm SU(2)}(x_1 \dots x_N)= \left[\min_{[P]} V_P(x_1 \dots x_N) \right]
\left( \alpha + 1 - \alpha P_{[P_{\rm min}]_0} \right)\;,
\label{minpotsun}
\end{equation}
where $P_{[P_{\rm min}]_0}$ denotes a color singlet state projector for
the minimum energy pairing and $\alpha$ is a large and positive constant.
For the pairing $12\;34\;56\;78\ldots$,
\begin{equation}
P_{[P_{\rm min}]_0}= |(12)_0\;(34)_0\;(56)_0\;(78)_0 \cdots
\rangle\langle (12)_0\;(34)_0\;(56)_0\;(78)_0 \cdots |\;.
\label{project}
\end{equation}
The potential choice in Eq.~(\ref{minpotsun})
 yields the required low-density limit
of color-singlet hadron clusters. The constant $\alpha$ must be large
relative to the maximum value of the spatial part of the potential, that is,
large relative to
$\left[\min_{[P]} V_P(x_1 \dots x_N) \right]$, in order for the
prescription to be sensible. The simulations are performed for quarks
in a box of length $L$ under (anti)periodic boundary conditions, so that the
spatial potential is bounded for fixed $L$.
The state $|(12)_0\;(34)_0\;(56)_0\;(78)_0 \cdots \rangle$ is
a manifest color singlet, though it is only a single component of
the full antisymmetric wavefunction for the quarks.
One must
begin, then, with a Slater determinant for the initial wavefunction
in order to guarantee antisymmetry under the exchange of any two
quarks. This is no different from the $U(1)$ calculation.
The $SU(2)$ prescription
given by Eqs.~(\ref{minpotsun}) and (\ref{project}) neglects
any contribution to the potential from paired triplet states which
are coupled to an overall singlet. The so-called ``hidden color''
forces are zero, then, in this prescription.

Computing observables with the color projector of Eq.~(\ref{project})
is cumbersome at best. In addition, the introduction of the constant
$\alpha$ is unappealing.
We shall choose, then, a different $SU(2)$
extension for use in this paper.
We shall retain the original $U(1)$ form of the potential,
Eq.~(\ref{minpot}), and
impose the needed constraints on our variational {\it Ansatz}.
The constraints we require are as follows.
\begin{itemize}
\item{%
The total wavefunction must be antisymmetric under quark exchange.
}
\item{%
The total wavefunction must be a color singlet.
}
\item{%
The total wavefunction must separate into color singlet
hadron clusters as pairs of quarks are pulled apart.
}
\end{itemize}
By imposing these constraints directly on the variational {\it Ansatz},
we have declared that the flux tube dynamics are such that
these properties are simply {\it guaranteed}.
For that reason, we call our extension to
$SU(2)_c$ the ``magic'' model. This model choice implies that
hidden color forces are now finite. The treatment of hidden color
here is compared with previous treatments of color in the quark exchange
model in
section~\ref{conferre}.

  We shall now proceed to construct the desired variational {\it Ansatz},
and shall illustrate the procedure to come by considering
the following {\it Ansatz} for four quarks:
\begin{eqnarray}
\Psi_{\rm trial}^{4q}(x_1,x_2,x_3,x_4) &=&
\exp[-\lambda(x_1-x_2)^2 - \lambda(x_3-x_4)^2] \;
f_{1234}^{\rm FG}\; |(12)_0 (34)_0\rangle
\nonumber\\
 &-&
\exp[-\lambda(x_1-x_3)^2 - \lambda(x_2-x_4)^2] \;
f_{1324}^{\rm FG} \;|(13)_0 (24)_0\rangle
\label{4qansatz}\\
 &+&
\exp[-\lambda(x_1-x_4)^2 - \lambda(x_2-x_3)^2] \;
f_{1423}^{\rm FG} \;|(14)_0 (23)_0\rangle \;.
\nonumber
\end{eqnarray}
The notation $|(12)_0\rangle$ denotes the pairing of quarks $1$ and $2$ into
a color singlet. The functions $f^{\rm FG}$ are the spatial wavefunctions
appropriate for a Fermi gas. Upon putting the quarks in a box of
length $L$ and choosing anti-periodic boundary conditions, we have
\begin{equation}
f_{1234}^{\rm FG} \equiv \cos {\pi \over L}(x_1 + x_2 - x_3 - x_4) \;.
\end{equation}
In $SU(2)_c$, two quarks can be put in a momentum state $+k$.
The variational parameter $\lambda$ describes the clustering of the
quarks into hadrons.
Equation (\ref{4qansatz}) has several important
properties.
\begin{itemize}
\item{%
 The wavefunction is explicitly antisymmetric under quark exchange.
}
\item{%
The $\lambda=0$ wavefunction is equivalent to a Fermi gas
Slater determinant.
}
\item{%
The wavefunction separates into {\it color singlet} hadron clusters
in the limit of large hadron-hadron separations, that is, for example,
in the $|x_1-x_2|, |x_3-x_4| \rightarrow\infty$ limit.
}
\end{itemize}
The above wavefunction {\it Ansatz} can be readily generalized to
$N$ quarks. One starts by generating all
distinct sets of quark pairs -- in the case of four quarks, there are three
such sets $(12),(34)$, $(13),(24)$, and $(14),(23)$. One then determines
the ``symmetry'' of each pairing with respect to one given pairing. That is,
one counts the number of permutations, $m$, required to bring the pairing
considered to the original pairing and then $(-1)^m$ determines whether
that pairing is even or odd. That sign is the over-all sign with which its
contribution is added to the wavefunction. The contribution that a particular
pairing $(12)(34)(56)(78)\cdots$ makes is
$\exp[-\lambda(x_{12}^2 +x_{34}^2 + x_{56}^2 + x_{78}^2)]
f^{\rm FG}(s_{12},s_{34},s_{56},s_{78}
\ldots)$, where $x_{ij}\equiv x_i - x_j$ and $s_{ij}\equiv x_i + x_j$. The
function $f^{\rm FG}$ is a symmetric function of its arguments, constructed
from cosines of the
$N/4$ momenta and all possible pairs of the $s_{ij}$'s.
The allowed momenta in the box are odd multiples of $\pi/L$, as we choose
to enforce anti-periodic boundary conditions. Thus, the function
$f^{\rm FG}$ for the configuration $12345678\ldots$ is
\begin{equation}
f^{\rm FG}_{12345678\ldots} =
\sum_{P_h}\;
{\rm  per} \left|_{ }
\begin{array}{lll}
\cos{\pi \over L}(P_h^1-P_h^2)  &\;\;\cos{\pi \over L}(P_h^3 - P_h^4)&
\cdots \\
\cos{3\pi \over L}(P_h^1-P_h^2) &\;\;\cos{3\pi \over L}(P_h^3 - P_h^4)&
\cdots \\
      \;\;\vdots                          &\;\;\;\;\vdots&  \;\;\vdots \\
 \cos({N \over 2} -1){\pi \over L}(P_h^1-P_h^2)
 &\;\;\cos({N \over 2} -1){\pi \over L}(P_h^3-P_h^4)& \cdots \\
\end{array}
\right|_{ } \;.
\label{fermidef}
\end{equation}
Note that ``per'' denotes a permanent.
A permanent is the symmetric version of
a determinant, so that the various terms are just added together.
In constructing the function $f^{\rm FG}_{12345678\ldots}$, we sum over all
possible sets of distinct pairs
of the hadronic coordinates $s_{ij}$, without regard to their order.
A set of hadronic coordinates is itself labelled ``$P_h$'', and there
are $(N/2)!/(2^{N/4}(N/4)!)$ different sets of distinct pairs for a given
$P_h$.
A member of $P_h$ is
denoted by $P_h^i$. For a given $P_h$,
we construct the permanent of the
allowed momenta and the $s_{ij}$ coordinates.
In writing Eq.~(\ref{fermidef}), we have assumed that the
number of quarks is a multiple of 4. It is only in this case that the
plane wave states for each free quark can be collapsed to an explicitly
real wavefunction.
Including the Fermi gas wavefunction is cumbersome, but it is absolutely
necessary. In its absence, the variational procedure is not well-posed,
as the $\lambda=0$ wavefunction is trivial.

In this manner,
the total
wavefunction can be constructed, and it
satisfies all the points
highlighted above for four quarks. Specifically, for eight quarks, we
have:
\begin{eqnarray}
\Psi_{\rm trial}^{8q}(x_1,\ldots,x_8) &=&
\exp[-\lambda(x_{12}^2 + x_{34}^2 + x_{56}^2 + x_{78}^2)] \;
f_{12345678}^{\rm FG}\; |(12)_0 (34)_0 (56)_0 (78)_0\rangle
\nonumber\\
 &-&
\exp[-\lambda(x_{13}^2 + x_{24}^2 + x_{56}^2 + x_{78}^2)] \;
f_{13245678}^{\rm FG}\; |(13)_0 (24)_0 (56)_0 (78)_0\rangle
\label{8qansatz}\\
 &+&
\exp[-\lambda(x_{14}^2 + x_{23}^2 + x_{56}^2 + x_{78}^2)] \;
f_{14235678}^{\rm FG}\; |(14)_0 (23)_0 (56)_0 (78)_0\rangle
\nonumber \\
&+& \ldots (102\; {\rm terms}) \;,
\nonumber
\end{eqnarray}
where
\begin{eqnarray}
f_{12345678}^{\rm FG}&=&\cos {\pi\over L}(s_{12} - s_{34})
\cos {3\pi\over L}(s_{56} - s_{78})
+ \cos {3\pi\over L}(s_{12} - s_{34})\cos {\pi\over L}(s_{56} - s_{78})
\nonumber \\
&+&
\cos {\pi\over L}(s_{12} - s_{56})\cos {3\pi\over L}(s_{34} - s_{78})
+ \cos {3\pi\over L}(s_{12} - s_{56})\cos {\pi\over L}(s_{34} - s_{78})
\\
&+& \cos {\pi\over L}(s_{12} - s_{78})\cos {3\pi\over L}(s_{34} - s_{56})
+ \cos {3\pi\over L}(s_{12} - s_{78})\cos {\pi\over L}(s_{34} - s_{56})
\;. \nonumber
\end{eqnarray}
The difficulty with this procedure
 is that the number of distinct pairs of quarks
grows greatly as $N$ increases. There are $N!/(2^{N/2}(N/2)!)$ terms in
the $N$ quark wavefunction,
so that the calculation quickly becomes unmanageable.
There are $105$ terms in the $8$ quark wavefunction, but there are
$10395$ terms in the $12$ quark wavefunction.

 Enumerating all the exchange terms by brute force is not
practical, nor is it necessary.  Our interest is not in the
wavefunction itself, but, rather, in the observables associated with
it. We are interested in the matrix elements
$\langle \Psi| \hat{O}|\Psi\rangle$, where $\hat O$ is a symmetric
-- and color-independent -- operator.
For example, $\hat{O}=\hat{T}=-{1\over 2}\sum_j \partial_j^2\;$; $\hat{O}$
can be the kinetic energy operator.
{\it It turns out that permutation symmetry greatly reduces the
number of non-trivial terms}.
Consequently, for 8 quarks, say, the
matrix element of the symmetric operator $\hat{O}$ can be written
\begin{eqnarray}
\langle \Psi_{\rm trial}^{8q} | &\hat{O}& |\Psi_{\rm trial}^{8q}\rangle
\propto \int\!\!\{dx_i\} f_{12345678}\; \hat{O} \;\Big[ (1)f_{12345678} -
(12) {1\over 2} f_{14325678}
\nonumber \\
&+& (32){1\over 4}f_{14365278} \quad + \quad (12){1\over 4}f_{14325876}
\quad -  \quad
(48){1\over 8}f_{14365872} \Big] \;,
\label{trickfa}
\end{eqnarray}
where
$f_{12345678}\equiv f_{12345678}^{\rm FG} \;\exp[-\lambda(x_{12}^2 +
x_{34}^2 + x_{56}^2 + x_{78}^2]$.
For the matrix elements of an operator $\hat{O}$ of the type above, one
need only count occurrences of topologically distinct terms. These are
terms which can not be brought into each other under rotation of the
coordinate indices. For example, for eight quarks, starting with
$12\;34\;56\;78$, there are
\begin{eqnarray}
\hbox{``singles''}  \leftrightarrow &14\;32\;56\;78&
\leftrightarrow \hbox{``[1]''} \nonumber \\
\hbox{``doubles''} \leftrightarrow &14\;36\;52\;78&
\leftrightarrow \hbox{``[2]''} \nonumber \\
\hbox{``triples''} \leftrightarrow &14\;36\;58\;72&
\leftrightarrow \hbox{``[3]''} \nonumber \\
\lefteqn{{\rm and}} \nonumber \\
\hbox{``disconnected doubles''} \leftrightarrow &14\;32\;58\;76&
\leftrightarrow \hbox{``[1+1]''}
\nonumber \nonumber \;.
\end{eqnarray}
The terms in brackets have been introduced as a convenient notation,
so that we can extend this construction to $N$ quarks.
We use ``singles'' or ``$[1]$'' to denote the single exchange of
quarks between two clusters, whereas ``doubles'' or ``$[2]$''
denotes the exchange of quarks between three clusters such that
one quark in the first cluster ends up in the third. This is
in contrast to a ``$[1+1]$'' exchange, in which two single exchanges
occur between two separate pairs of clusters.
In Eq.~(\ref{trickfa}), the numbers in parentheses
are the number of terms of that type
in the original wavefunction, so that their sum is
105. The fractions are the color matrix
elements for terms of that type, note that
$\langle (12)_0 (34)_0 (56)_0 \ldots | (12)_0 (34)_0 (56)_0 \ldots \rangle
= 2^{N/2}$. Since we do compute the color matrix element, we need only
consider permutations between one member of a particular pair of quarks;
in the above, for example, we have kept the odd quarks fixed and
considered exchanges of the even ones.
The sum of all the
coefficients, then, including the color matrix element, is $(N/2)!$, or 24
for the eight quark case.

For $N$ quarks, one can write the general form of the
coefficient, given its topological structure. Moreover, one can enumerate
all the possible topological structures for a fixed number of quarks.
In order to discuss this, we shall use the ``$[i+j+k+\cdots]$'' notation
defined above.

The distinct topological terms for a fixed number of quarks are given by
$[i+j+k+\cdots]$, where the range of $i,j,k,\ldots$ satisfies the
following bounds.
That is, $i$ ranges from $1$ to the upper
bound $\xi=N/2 -1$ and $j,k,\ldots$
range from $0$ to $\xi$, subject to the constraints that
$i+j+k+ \cdots
(\kappa\;{\rm terms}) \le \xi - \kappa +1$
and $i\le j\le k\le \ldots$ when $j,k,\ldots\ne 0$.
For eight quarks, the terms are
$[1],\;[2],\;[3],\;[1+1]$. For twelve quarks, the terms are
$[1],\;[2],\;[3],\;[4],\;[5],\;[1+1],\;[1+2],\;[1+3],\;[2+2],\;[1+1+1]$.
The formula for the overall coefficient of the $[i+j+k+\cdots]$ term can
be expressed in closed form. That is,
\begin{equation}
C_{[i+j+k+\cdots]} = {1\over s}
\left[{(-1)^{i+j+k+ \cdots} \over {(i+1)(j+1)(k+1)\cdots}}
\right](n_H)\cdots(n_H-i-j-k) \;,
\end{equation}
where $n_H\equiv N/2$, and $s$ is a symmetry factor. $s$ is determined
by the number of repeated exchanges. That is, if $[i+j+k+\cdots]$ has
$I$ terms of type $i$ and $J$ terms of type $j$, and all
the other terms are distinct, then $s=I! \;J!$. For example,
with $[1+1]$, $s=2$.
Thus, we have a strategy for the computation of
the $\langle \Psi_{\rm trial}^{Nq} | \hat{O} | \Psi_{\rm trial}^{Nq}\rangle$
matrix element:
\begin{equation}
\langle \Psi_{\rm trial}^{Nq} | \hat{O} | \Psi_{\rm trial}^{Nq}\rangle
\propto \int \!\! \{dx_i\} f_{1234\cdots N} O \Psi_{\rm trick}^{Nq}
\label{trickpsi}
\end{equation}
where
\begin{equation}
\Psi_{\rm trick}^{Nq}=f_{1234\cdots N} +
\sum_{i,j,k,\ldots}\!\!\! { }^{'}
C_{[i+j+k+\cdots]} f_{[i+j+k+\cdots]} \;.
\end{equation}
The notation $\sum'$ means that the sum is restricted.
As above, we require that
$i\le j\le k \ldots$ when $j,k,\ldots\ne 0$ and
$i+j+k+ \cdots (\kappa\;{\rm terms}) \le N/2 - \kappa $.
Note that $i$ starts at one, whereas $j,k,\ldots$ begin at zero.
In Sec.~{\ref{monte}}, we discuss the stochastic evaluation of
Eq.~(\ref{trickpsi}) and the resulting computation of
the total energy and the two-body correlation function
with density.

We can compute
the $\langle \Psi_{\rm trial}^{Nq} | \hat{O} | \Psi_{\rm trial}^{Nq}\rangle$
matrix element in another manner. That is,
\begin{equation}
\langle \Psi_{\rm trial}^{Nq} | \hat{O} | \Psi_{\rm trial}^{Nq}\rangle
\propto \int\!\! \{dx_i\}
\left( \Psi_{\rm trial}^{Nq}\right)_{\uparrow\downarrow\uparrow\downarrow
\ldots}
O
\left( \Psi_{\rm trial}^{Nq}\right)_{\uparrow\downarrow\uparrow\downarrow
\ldots}
\;.
\label{tricksqua}
\end{equation}
Here ``$\uparrow$'' and ``$\downarrow$'' denote the two color states
of the model, and
$( \Psi_{\rm trial}^{Nq})_{\uparrow\downarrow\uparrow\downarrow\ldots}$
denotes a projection of the full wavefunction on the
$\langle \uparrow \downarrow\uparrow \downarrow\ldots|$ state.
We have written the integrand in
Eq.~(\ref{tricksqua}) as a manifestly positive definite
quantity. This is highly convenient for a stochastic evaluation of
the integral, though it is at the cost of a more complicated wavefunction.
This wavefunction contains $(N/2)!$ terms, but it is simply related
to $\Psi_{\rm trick}^{Nq}$. Starting with $\Psi_{\rm trick}^{Nq}$, we
enumerate all the terms of a particular topological class. Thus, for
8 quarks, in place of the term with
\begin{equation}
 [1] \leftrightarrow 14\; 32\; 56\; 78
\end{equation}
we explicitly include all 6 terms of that type. That is,
\begin{eqnarray}
 14\; & \;32\;  \;56\;  \;78 \nonumber\\
 16\; & \;34\;  \;52\;  \;78 \nonumber\\
 18\; & \;34\;  \;56\;  \;72 \nonumber\\
 12\; & \;36\;  \;54\;  \;78 \nonumber\\
 12\; & \;38\;  \;56\;  \;74 \nonumber\\
 12\; & \;34\;  \;58\;  \;76 \nonumber
\end{eqnarray}
In this manner,
$( \Psi_{\rm trial}^{Nq})_{\uparrow\downarrow\uparrow\downarrow\ldots}$
can be readily generated.

We shall study the benefits and disadvantages
offered by Eq.~(\ref{trickpsi})
and Eq.~(\ref{tricksqua}) in Sec.~{\ref{monte}}.


\section{COMPARISON WITH OTHER MODELS}
\label{conferre}

Here we compare the model defined by Eq.~(\ref{4qansatz})
with previous attempts to incorporate $SU(N)$ color.
To our knowledge, our model study is the first complete
treatment of $SU(2)_c$ in quark matter. To make our
treatment of hidden color forces clear, we first consider
the original $SU(N)_c$ study of
Lenz {\it et al.} \cite{lenz} in the
$q^2 \overline{q}^2$ system.
For a particular pairing $(12)(34)$, for example, they postulate
that confining forces of differing range may reside in the orthogonal sectors
$|(12)_0(34)_0\rangle$ and $|(12)_1(34)_1\rangle_0$. The
sector in which the individual quarks are paired to triplets before
being paired to an overall singlet is the ``hidden color'' sector.

The color kets of Eq.~(\ref{4qansatz}) can be recast into the
framework described above. That is, the kets
$|(13)_0(24)_0\rangle$ and $|(14)_0(23)_0\rangle$ can be expressed as linear
combinations of $|(12)_0(34)_0\rangle$ and $|(12)_1(34)_1\rangle_0$.
Following Lenz {\it et al.} \cite{lenz}, we
introduce the coordinates
\begin{eqnarray}
x &=& {1\over 2}(x_1 + x_3 - x_2 - x_4) \nonumber \\
y &=& {1\over 2}(x_1 + x_2 - x_3 - x_4)
\label{lenzxyz}\\
z &=& {1\over 2}(x_1 + x_4 - x_3 - x_2) \;.\nonumber
\end{eqnarray}
These are useful reduced variables, as they describe the relative separation
of hadrons with a particular quark content. With Eq.~(\ref{lenzxyz}),
we can rewrite Eq.~(\ref{4qansatz}) as
\begin{eqnarray}
\Psi^{4q}_{\rm trial}&=&
\Bigg[
\exp[-2\lambda (x^2 + z^2)]\;f_{1234}^{\rm FG}
-{1 \over 2}\exp[-2\lambda (y^2 + z^2)]\;f_{1234}^{\rm FG} \nonumber \\
&{ }&-{1 \over 2}\exp[-2\lambda (x^2 + y^2)]\;f_{1423}^{\rm FG}
\Bigg]\;|(12)_0(34)_0\rangle
\nonumber \\
&{ }&+
\Bigg[
-{1 \over 2}\exp[-2\lambda (y^2 + z^2)]\;f_{1234}^{\rm FG}
+{1 \over 2}\exp[-2\lambda (x^2 + y^2)]\;f_{1423}^{\rm FG}
\Bigg]\;|(12)_1(34)_1\rangle_0
\label{4qhidden}
\end{eqnarray}
where  $ |(12)_1(34)_1\rangle_0 \equiv
|(12)\;11\rangle |(34)\;\hbox{1-1}\rangle
+ |(12)\;\hbox{1-1}\rangle|(34)\;11\rangle -
|(12)\;10\rangle|(34)\;10\rangle$. Note that in $|(12)\;{\rm IJ}\rangle$
$I$ and $J$ denote the $z$ component of the color of quarks
1 and 2, respectively.
We see manifestly, then, that hidden color states are equivalent
to the coupling of the other pairs of quarks to singlets, and, thus,
are trivial. Indeed, in
the absence of active gluon degrees of freedom, this is generally
so.
In the channel with the pairing $(12)(34)$, the relative separation
of $(12)$ and $(34)$ is determined by $y$ -- so it is the ``channel''
variable. The exchange
terms in the color singlet and hidden color sectors have the same behavior
in $y$, so that the spatial range of the confining
forces in the two sectors is the same.
Lenz {\it et al.} \cite{lenz} considered the $q^2\overline q^2$
system as a function of a parameter specifying the relative
spatial range of the color singlet (comprised of color singlets)
and hidden color sectors. Our model of hidden color
forces maps to one particular case in their study, although there is
no especial physics reason for our choice.

The full $SU(2)$ treatment of Sec.~{\ref{themagmod}} can also
be compared with the approximate treatment of Horowitz and
Piekarewicz \cite{cjhjp3d}. Their work was in $SU(3)$ color and three
spatial dimensions; we now consider this approach in
$SU(2)_c$ for purposes of comparison.
In their treatment of color, the quarks are given {\it fixed} colors.
(Watson \cite{watson} has also used this starting point.)
For this reason, in the discussions to come, we will refer to this
approach as the ``painted'' model. In constrast to Eq.~(\ref{minpot}),
the potential is no longer color-blind, but, rather, consists of
the minimum energy pairing of quarks of differing color.
The construction
of a totally antisymmetric, color singlet variational Ansatz is
straightforward. That is,
\begin{equation}
 \Psi^{Nq}_{\rm trial}
=\exp[-\lambda \hat{V}_{\rm RB}]\;
 \Psi^{\rm FG}_{\rm R} \Psi^{\rm FG}_{\rm B} \;,
\label{paintans}
\end{equation}
where
\begin{equation}
\hat{V}_{\rm RB} = \min_{[L,M]} \sum_{i,j=1}^{N/2} v(x^r_{L^{i}} - x^b_{M^{j}})
\end{equation}
such that $v(x^r_{L^1} - x^b_{M^1})=(x^r_{L^1} - x^b_{M^1})^2/2$.
The $L$ and $M$ are elements of the permutation group $S_{(N/2)}$, and
$L^i$ is the $i^{\rm th}$ member of the element
$L$. We have presumed that there are equal numbers of red (R) and
blue (B) quarks and that their coordinates are labelled by the appropriate
superscripts. Note that
$\Psi^{\rm FG}_{\rm R}$ is the Slater determinant for the red quarks.
The following convenient form
is equivalent to a Slater determinant:
\begin{equation}
\Psi^{\rm FG}_{\rm R}=\prod_{n,m \;{\rm odd}}^N
\sin\left[ {\pi \over L} (x_n -x_m) \right] \;.
\end{equation}
This method is simple to implement and is able to treat many quarks
with ease. We will compare the full $SU(2)_c$ results to the
simple approximation described here for two reasons. First, we wish to
examine the extent to which the predictions of the painted model are
robust. Its primary advantage is convenience; we
can not expect the painted model to be exact as all color flipping
transitions, {\it i.e.},
${\rm RB} \leftrightarrow {\rm BR}$, are neglected. However,
the differences might not be large. Second, we hope that any departures
between the full $SU(2)_c$ and painted calculations will give us greater
insight into the nature of the correlations in the full calculation.

We can begin to understand the difference between the painted (approximate
$SU(2)_c$) and magic (full $SU(2)_c$)
models through consideration of their variational {\it Ans\"atze}
for four quarks. That is, for the magic model,
\begin{eqnarray}
\langle \uparrow\downarrow\uparrow\downarrow | \Psi^{4q}_{\rm M}\rangle
=&\exp\left[ -\lambda [(x_1 -x_2)^2 + (x_3 -x_4)^2]
\right]\cos{\pi \over L}(x_1+x_2-x_3-x_4)
\nonumber \\
&-
\exp\left[ -\lambda [(x_1 -x_4)^2 + (x_2 -x_3)^2]
\right]\cos{\pi \over L}(x_1+x_4-x_2-x_3)
\label{4qmagiccf}
\end{eqnarray}
whereas, in the painted model,
\begin{equation}
\langle \uparrow\downarrow\uparrow\downarrow | \Psi^{4q}_{\rm P}\rangle
=\exp\left[
-\lambda \min[V_{(12),(34)},V_{(14),(23)}]
\right]\sin{\pi\over L}(x_1-x_3) \sin{\pi\over L}(x_2-x_4)
\label{4qpaintcf}
\end{equation}
where quarks $1,3$ are red (``up'') and quarks $2,4$ are blue (``down'').
The Hamiltonians
of the two models differ, but it is still useful to compare the structure
of the {\it Ans\"atze} as two quarks of the same color approach each
other. In this way, we can understand more about the dynamical content
of the two models. Letting $x_1-x_3\equiv \varepsilon$ be small, we
have
\begin{eqnarray}
\langle \uparrow\downarrow\uparrow\downarrow | \Psi^{4q}_{\rm M}\rangle
=&\exp\left[ -\lambda [(x_2 -x_3)^2 + (x_3 -x_4)^2]
\right]
\{ -2\varepsilon{\pi \over L} \sin{\pi \over L}(x_2-x_4)
 \nonumber \\
&+ 2 \lambda(x_2 -x_4)\varepsilon \cos{\pi \over L}(x_2-x_4)
+ O(\varepsilon^2) \}
\label{x13magic}
\end{eqnarray}
for the full $SU(2)_c$ calculation and
\begin{equation}
\langle \uparrow\downarrow\uparrow\downarrow | \Psi^{4q}_{\rm P}\rangle
=\exp\left[ -\lambda [(x_2 -x_3)^2 + (x_3 -x_4)^2]
\right] \{ \varepsilon{\pi \over L}\sin{\pi \over L}(x_2-x_4)
+ O(\varepsilon^2)\}
\label{x13paint}
\end{equation}
for the painted model. Equation (\ref{x13magic}) contains an
additional term in leading order in $\varepsilon$. This term
does not have the same manifest $L$ dependence as the term which
both Eqs.~(\ref{x13magic}) and (\ref{x13paint}) share. Indeed, this
additional term suggests that the Fermi ``wound'' should heal
more quickly in the full $SU(2)_c$ calculation than in the painted
case, even for large $L$. This qualitative difference in the
{\it Ans\"atze}'s structure will be manifest in a comparison of
the models' two-body densities.


\section{MONTE CARLO COMPUTATIONS}
\label{monte}

In this section, we shall discuss the techniques necessary to evaluate
the integrals in Eq.~(\ref{trickpsi}) and Eq.~(\ref{tricksqua}),
so that
the total energy
of the system and its two-body density
may be computed. First, we shall
discuss the computation of the system's energy. Let us consider, then,
the evaluation of the kinetic energy: $T=-({1/2})\sum_i \partial_i^2$.
Both Eq.~(\ref{trickpsi}) and Eq.~(\ref{tricksqua}) have
terms purely
of the form
\begin{equation}
f^{\lambda}_{12345678\ldots}f^{\rm FG}_{12345678\ldots} \;,
\end{equation}
where
\begin{equation}
f^{\lambda}_{12345678\ldots}=\exp[-\lambda[(x_1-x_2)^2 + (x_3-x_4)^2
+ (x_5-x_6)^2 + (x_7-x_8)^2 + \cdots]]
\end{equation}
and $f_{12345678\ldots}^{\rm FG}$ has been introduced in Eq.~(\ref{fermidef}).
For wavefunctions of this structure, the kinetic energy can be calculated
in a straightforward way. Note that
\begin{mathletters}
\begin{equation}
\partial_1 f_{1234\ldots}^{\rm FG}=\partial_2 f_{1234\ldots}^{\rm FG}
\end{equation}
whereas
\begin{equation}
\partial_1 f_{1234\ldots}^{\lambda}=-\partial_2 f_{1234\ldots}^{\lambda} \;,
\end{equation}
\end{mathletters}
so that $\sum_{i=1}^2 \partial_i f^{\lambda}_{1234\ldots}
\partial_i f^{\rm FG}_{1234\ldots}=0$. This cancellation occurs pairwise,
so that
\begin{mathletters}
\begin{eqnarray}
  -{1\over 2}\sum_{j=1}^N
\partial_j^2 (f_{1234\ldots}^{\lambda} f_{1234\ldots}^{\rm FG})
&=& -{1\over 2}
\sum_{j=1}^N [(\partial_j^2 f_{1234\ldots}^{\lambda})f_{1234\ldots}^{\rm FG}
+ f_{1234\ldots}^{\lambda}(\partial_j^2 f_{1234\ldots}^{\rm FG})]
\label{derivcalc:a}
\\
&=& [ \lambda N - 4 \lambda^2 ((x_1-x_2)^2 + (x_3-x_4)^2 + \cdots)) ]
f_{1234\ldots}^{\lambda} f_{1234\ldots}^{\rm FG}
\label{derivcalc:b}
\\
&{ }&+ t_{\rm FG}N f_{1234\ldots}^{\lambda} f_{1234\ldots}^{\rm FG} \nonumber
\end{eqnarray}
\end{mathletters}
Note that $t_{\rm FG}$ is the Fermi gas kinetic energy per quark
for a finite number
of quarks. Upon choosing anti-periodic boundary conditions, the
allowed momenta in the box are odd integers, and
\begin{equation}
t_{\rm FG}^{\rm a-p} = {2\pi^2 \over {N L^2}}\sum_{i=1}^{N/4} (2i-1)^2 \;,
\end{equation}
so that
\begin{equation}
t_{\rm FG}^{\rm a-p} = {\pi^2 \rho^2 \over 24}(1 - {4 \over N^2}) \;,
\label{antifermi}
\end{equation}
where the density $\rho=N/L$ has been introduced. For large N, this
is the continuum $SU(2)$ Fermi gas kinetic energy; the corrections to
this limit are of order $1/N^2$. We hope
that this effect will persist for other observables and for finite
$\lambda$.

 With the use of Eq.~(\ref{derivcalc:b}) and Eq.~(\ref{antifermi}), the form of
the derivative terms can be realized analytically. We are able, then, to write
down the precise form of the matrix elements we wish to calculate.
Presuming that the initial matrix element is of the
form given in Eq.~(\ref{tricksqua}) and
letting
$\tilde{A} \equiv
(\Psi_{\rm trial}^{Nq})_{\uparrow\downarrow\uparrow\downarrow\ldots}$,
the kinetic energy per particle $\langle t \rangle_{\lambda}$ and
the potential energy per particle $\langle v \rangle_{\lambda}$ become
\begin{equation}
 \langle t \rangle_{\lambda} = t_{\rm FG} + \lambda -
{4\lambda^2 \over N}{ \int\{dx_i\}\tilde{A}\tilde{g}
\over{\int\{dx_i\} \tilde{A}^2  }}
\label{kelambda}
\end{equation}
and
\begin{equation}
 \langle v \rangle_{\lambda} =
{1 \over N}{ \int\{dx_i\}\tilde{A}V\tilde{A}
\over{\int\{dx_i\} \tilde{A}^2  }} \;,
\label{potlambda}
\end{equation}
where
\begin{eqnarray}
\tilde{g}&\equiv&
[(x_1 - x_2)^2 + (x_3 - x_4)^2 + (x_5 - x_6)^2 +
(x_7-x_8)^2 + \cdots]f_{12345678\ldots}  \nonumber \\
&{ }&-
[(x_1 - x_4)^2 + (x_3 - x_2)^2 + (x_5 - x_6)^2 +
(x_7-x_8)^2 + \cdots]f_{14325678\ldots}  \nonumber \\
&{ }&+\; ((N/2)! -2) \;{\rm terms}
\label{tilgdef}
\end{eqnarray}
and $V$ denotes the usual many-body potential,
Eq.~(\ref{minpot}).
Note that the total energy per particle is
$\langle \varepsilon \rangle_{\lambda} = \langle t \rangle_{\lambda} +
\langle v \rangle_{\lambda}$.
The numerator of Eq.~(\ref{kelambda}) no longer has a
$\tilde{A}^2$ structure; indeed we ``measure'' $\tilde{g}/\tilde{A}$ with
respect to the distribution $\tilde{A}^2$.
Note that configurations with $\tilde{A}=0$ are generated with zero weight,
so that the procedure we effect actually works
rather well.

Before we proceed, we should mention how the potential, Eq.~(\ref{minpot}),
is calculated. The economics literature \cite{econo} abounds with
numerical algorithms for minimization problems -- we need only recall
the ``travelling salesman'' problem. For our purposes, determining the
optimum pairing in one spatial dimension is simple. The continuous system
is modelled by quarks on a ring; a particular
quark on that ring can be paired with its neighbor to its left
or its right, and that one choice determines the partners for all the
other quarks on the ring. Thus, we only have to calculate the energy
of the two different arrangements to determine the optimum pairing.
Clearly, any arrangement which involves overlapping strings can not
be an optimum pairing.

Our results depend on the variational
parameter $\lambda$, and we wish to determine the optimal $\lambda$
in an efficient way.
We must find $\lambda_t$ such that
\begin{equation}
{{\rm d}\over {{\rm d}\lambda}}
\langle \varepsilon \rangle_{\lambda}\Bigg|_{\lambda=\lambda_t}
= 0 \;.
\end{equation}
We can do this by exploiting how the
matrix elements scale in $\lambda$ and $L$. The wavefunction {\it Ansatz}
itself is a function of the dimensionless variable $\lambda L^2$ alone,
but the final matrix elements do have a manifest $L$ dependence.
That is,
\begin{mathletters}
\begin{eqnarray}
\langle t \rangle_{\lambda} &\sim& \Big[ {1\over L^2} \Big] \\
\langle v \rangle_{\lambda} &\sim& \Big[ L^2 \Big] \;.
\end{eqnarray}
\end{mathletters}
The $t$ and $v$ matrix elements' dependence on $L$ differ, so that
the optimal $L$ -- and, hence, the optimal $\lambda$ --
can be computed directly from $d\langle t \rangle_{\lambda}/d\lambda$
and $d\langle v \rangle_{\lambda}/d\lambda$. In this way, a single calculation
for fixed $\lambda L^2$ suffices to determine the optimal variational
parameter and the observables for that $\lambda$.
We calculate $d\langle t \rangle_{\lambda}/d\lambda$
and $d\langle v \rangle_{\lambda}/d\lambda$
directly, as
computing the numerical derivative in $\lambda$
of the stochastically calculated $t$ and $v$ matrix
elements is unwieldly. Moreover, sampling the appropriate derivatives
directly does not require much additional effort. Thus, we compute
\begin{eqnarray}
 {d \over d\lambda}\langle t \rangle_{\lambda} &=& 1 -
{8\lambda \over N}{ \int\{dx_i\}\tilde{A}\tilde{g}
\over{\int\{dx_i\} \tilde{A}^2  }}
+
{4\lambda^2 \over N}{ \int\{dx_i\}(\tilde{A}\tilde{h} + \tilde{g}^2)
\over{\int\{dx_i\} \tilde{A}^2  }} \nonumber \\
&-&
{8\lambda^2 \over N}\left[{ \int\{dx_i\}\tilde{A}\tilde{g}
\over{\int\{dx_i\} \tilde{A}^2  }} \right]^2 \;,
\label{dkelambda}
\end{eqnarray}
and
\begin{equation}
 {d \over d\lambda}\langle v \rangle_{\lambda} =
-{2 \over N}{ \int\{dx_i\}\tilde{A}V\tilde{g}
\over{\int\{dx_i\} \tilde{A}^2  }}
+
{2 \over N}{ \int\{dx_i\} \tilde{A}\tilde{g}
\int\{dx_i\}\tilde{A}V\tilde{A}
\over{\left[ \int\{dx_i\} \tilde{A}^2 \right]^2 }} \;,
\label{dpotlambda}
\end{equation}
where $\tilde{g}$ is given by Eq.~(\ref{tilgdef}) and
\begin{eqnarray}
\tilde{h}&\equiv&
[(x_1 - x_2)^2 + (x_3 - x_4)^2 + (x_5 - x_6)^2 +
(x_7-x_8)^2 + \cdots]^2 f_{12345678\ldots}  \nonumber \\
&{ }&-
[(x_1 - x_4)^2 + (x_3 - x_2)^2 + (x_5 - x_6)^2 +
(x_7-x_8)^2 + \cdots]^2f_{14325678\ldots}  \nonumber \\
&{ }&+\; ((N/2)! -2) \;{\rm terms} \;.
\label{tilhdef}
\end{eqnarray}
As in Eq.~(\ref{kelambda}), we measure $\tilde{g}/\tilde{A}$, as
well as $\tilde{h}/\tilde{A}$ and $(\tilde{g}/\tilde{A})^2$,
with respect to the distribution $\tilde{A}^2$.
We can now proceed to construct the optimal variational solution
for some fixed $\lambda L^2$. We shall label the observables computed
at some $\rho_0$ and $\lambda_0$ by ``$0$'', and the observables
at the desired scaled point by ``$t$''. That is, we want $\rho_t$
and $\lambda_t$ such that
\begin{equation}
{ d \langle t \rangle_t \over { d\lambda}}
+ { d \langle v \rangle_t \over { d\lambda}} =0 \;,
\end{equation}
so that
\begin{equation}
{ d \langle t \rangle_0 \over { d\lambda}}
+ \left[{\rho_0 \over \rho_t}\right]^4
{ d \langle v \rangle_0 \over { d\lambda}} =0 \;.
\end{equation}
Thus,
the optimal $L$ -- or, rather, $\rho_t$ --
for fixed $\lambda L^2$ is given
by
\begin{equation}
\rho_t = \rho_0 \left[ -
{{d\langle t \rangle_0 \over { d\lambda}}
\over
{ {d\langle v \rangle_0 \over {d\lambda}}}}
\right]^{-1/4} \;.
\end{equation}
Introducing
\begin{equation}
\eta = \left[{ \rho_0 \over \rho_t }\right]^2 \;,
\end{equation}
we can compute the optimum $\lambda$:
\begin{equation}
\lambda_t = {\lambda_0 \over \eta} \;,
\end{equation}
as well as the kinetic and potential energies at the new point:
\begin{equation}
\langle t \rangle_t = {{\langle t \rangle_0} \over \eta} \;,
\end{equation}
and
\begin{equation}
\langle v \rangle_t = \eta{\langle v \rangle_0}  \;.
\end{equation}
Of course, $\langle \varepsilon \rangle_t = \langle t \rangle_t +
\langle v \rangle_t$. These scaling relations serve to generate the
optimal variational solution for some fixed $\lambda L^2$. Their
utility depends on the
energy having a smooth, parabolic behavior in the region of $\lambda_t$,
$\rho_t$. Even if
this constraint is satisfied, as it is here, we can not strictly
say that we have determined the best variational solution
for a fixed density. For a fixed density, there exist two
separate parametric families -- those with finite $\lambda$
and those with $\lambda=0$. Scaling relations also exist for the
$\lambda=0$ family.
The $\lambda=0$ solution may have a
lower energy than the finite $\lambda$ solution for some fixed density;
the relative energies of the two families must be determined ``empirically''.

 In principle, we have two distinct strategies,
Eqs.~(\ref{trickpsi}) and (\ref{tricksqua}), for computing the
observables associated with the Hamiltonian of Eqs.~(\ref{hamdef}) and
(\ref{minpot}). The
wavefunction in Eq.~(\ref{trickpsi}) is much simpler, but
we must compute its normalization explicitly. We have compared the
two schemes for both 8 and 12 quarks, and have found Eq.~(\ref{tricksqua})
to be the method of choice. The two strategies do yield identical results
within error bars, as they should, but {\it many} more integration
points are required to yield good results in the case where we must
calculate the normalization. Consequently, the formulae and results we
present exploit Eq.~(\ref{tricksqua}).

Thus far, we have described the procedure by which the optimal
variational calculation of the energy is effected. However, we are
interested in more than the energetics of the system; we
would also like to study the structure of the ground state.
We are particularly interested in the correlations induced in the
ground state as a result of the internal color degree of freedom and
global antisymmetry. Thus, we shall now describe how the two-body
density may be calculated as well. The two-body density operator,
$\hat\rho_2(r)$,
is defined as
\begin{equation}
\hat\rho_2(r) = {1 \over L} \sum_{i\ne j} \delta(r - (x_i - x_j)) \;.
\label{rho2def}
\end{equation}
This operator is manifestly symmetric, so that we may compute the
resulting matrix element in the manner indicated by Eq.~(\ref{tricksqua}).
We choose to normalize the matrix element of the two-body density
by the square of the one-body density; that is, we shall compute
the matrix element of $\hat{\rho}_2(r)/\rho_0^2$. With
this choice of normalization, the low-density limit of this matrix element is
\begin{equation}
{ \langle \rho_2(r) \rangle_{\rho_0 \rightarrow 0}
\over { \rho_0^2}} = {|\Psi_{\rm isol}(r)|^2 \over \rho_0} \;.
\label{rho2low}
\end{equation}
$|\Psi_{\rm isol}(r)|^2$ is
the square of the isolated hadron wavefunction; here
\begin{equation}
|\Psi_{\rm isol} (r)|^2 = \left({1\over{{\sqrt 2}\pi}} \right)^{1/2}
e^{- {\sqrt{2}\over 2}r^2} \;,
\label{psiisol}
\end{equation}
so that the low density limit of the clustering parameter $\lambda$ is
\begin{equation}
\lambda(\rho\rightarrow 0)= {1\over {2\sqrt{2}}} \simeq .35 \;.
\label{lowrholamb}
\end{equation}
The matrix element of Eq.~(\ref{rho2def}) measures the correlation
of any two quarks as a function of their separation $r$
in the system's ground state.
At low densities, the quark-quark correlation function is
controlled by the isolated hadron wavefunction, as shown in
Eq.~(\ref{rho2low}). At high densities, the system becomes a non-relativistic
Fermi gas. In this limit, the two-body density is
\begin{equation}
{ \langle \rho_2(r) \rangle_{\rm FG}
\over { \rho_0^2}} = 1 - {1\over 2} j_0^2(k_F r) \;,
\label{rho2high}
\end{equation}
where, in $SU(2)_c$, the Fermi momentum $k_F$ is given by
$k_F={\pi \rho_0 /2 }$. There is a 50\% probability that any two quarks
in $SU(2)_c$ will have different color states; hence, the two-body
density at $r=0$ is $1/2$.

The Metropolis Monte Carlo procedure we use allows us to calculate
the two-body density in a straightforward way. That is, we generate the
quarks'
coordinates according to the distribution
$|( \Psi_{\rm trial}^{Nq})_{\uparrow\downarrow\uparrow\downarrow\ldots}|^2$,
in order
to calculate matrix elements according to
Eq.~(\ref{tricksqua}). We can compute the matrix element of
Eq.~(\ref{rho2def}) with $r$
by simply tabulating the occurrences of $x_i - x_j$ in the generated
distribution over the interval $[0,L]$ in bins of $\Delta r$.
The resulting histogram is a discrete representation of the
two-body density with $r$. As in the case with the energy, scaling
can be used to calculate the two-body density
for the optimum value of the variational parameter at some fixed density
in an economical manner. The quantity
${ \langle \rho_2(r) \rangle_{\rho_0 \rightarrow 0} / { \rho_0^2}}$
is itself
$L$-independent, but the scale in $r$ over which it is plotted
does depend on $L$.  The step-size in $r$ upon scaling from
$\rho_0$ to $\rho_t$ transforms according to
\begin{equation}
\left(\Delta r\right)_t =\left(\Delta r\right)_0 \left(
{\rho_0 \over \rho_t} \right) \;.
\end{equation}
In this manner, the two-body density at the scaled values of
the density and variational parameter can be computed from some
initial $\rho_0$ and $\lambda_0$.


\section{RESULTS}
\label{results}

In this section, we shall present our variational Monte Carlo results,
computed using Eq.~(\ref{tricksqua}).
We have computed the energy and correlation
function as a function of density for eight and twelve quarks.
The low and high density limits of the
model are known exactly, so that it is useful to review those results
before proceeding to a discussion of the simulations. At low densities,
the energy per quark $\varepsilon$
and the variational parameter $\lambda$ are given by
the isolated hadron values. The isolated hadron wavefunction in our
units is given by Eq.~(\ref{psiisol}), so that at low densities,
\begin{mathletters}
\begin{eqnarray}
\langle\varepsilon\rangle_{\lambda} &\rightarrow& {1 \over {2\sqrt{2}}}
\simeq .35 \\
\langle t\rangle_\lambda &\rightarrow& \langle v \rangle_\lambda
\rightarrow {\langle \varepsilon\rangle_\lambda \over 2} \simeq .18 \;.
\label{lowrhoe:b}
\end{eqnarray}
\end{mathletters}
Equation (\ref{lowrhoe:b}) is a consequence of the virial theorem in the
isolated hadron limit. The isolated hadron variational parameter is
given in Eq.~(\ref{lowrholamb}).

  We shall first examine the energy and variational parameter as a function
of density before turning to the correlation function results.

\subsection{ENERGY VS. DENSITY}

  Figure {\ref{fig8qfig1}} shows the variational parameter $\lambda$ as
a function of density for eight quarks. The indicated points are the
lowest energy solutions as a function of density
 for the finite $\lambda$ parametric family. At low density, the scaling
solution approaches the isolated hadron limit value, Eq.~(\ref{lowrholamb}).
At high density, the energies of the optimal solutions of
the finite $\lambda$
and zero $\lambda$ families cross; the variational
parameter drops to zero for $\rho \sim 1$. The manner in which this
level crossing occurs is shown in Fig.~\ref{fig8qfig2}.
At low density, the energy as a function of the variational
parameter has a single extremum. As the density increases, a double-welled
structure develops; at sufficiently high density, the zero $\lambda$
solution is lower in energy. The two extrema are well-separated in
$\lambda$; this generates the sudden drop in $\lambda$ with
$\rho$ seen in Fig.~\ref{fig8qfig1}.
This behavior is a non-trivial dynamical consequence of the model.

  The meaning of ``high'' density is not yet clear,
as the scale in Fig.~\ref{fig8qfig1} is in dimensionless units. In the units
 of this paper, the rms separation of the quarks in
an isolated ground state hadron (noting Eqs.~(\ref{psiisol})
and (\ref{lowrholamb})) is
\begin{equation}
\langle r^2 \rangle^{1\over 2} = {1\over{ 2\sqrt{\lambda_{\rm isol}}}}
\simeq .84
\end{equation}
so that one hadron in an isolated hadron length implies a quark
number density of
2.4. The number density of realistic nuclear matter at saturation
is $.17\; {\rm fm}^{-3}$. The charge radius of the proton in
the non-relativistic quark model \cite{isgurkarl} is
$.5\;{\rm fm}$, so that putting one proton in the volume given by
that radius corresponds to a number density of $1.9\;{\rm fm}^{-3}$.
Consequently, the density .95 in our units corresponds roughly
to 4 times nuclear matter density, although one must be cautious in
assigning dimensions in this simple model.

\vbox to 3.75in{\vss\hbox to 5.625in{\hss
{\includegraphics{8qfig1.ps}}\hss}}
\nobreak
{\noindent\narrower{{\bf FIG.~\protect{\ref{fig8qfig1}}}.
The variational parameter $\lambda$ as a function
of $\rho$ for eight quarks. The points indicate the optimum finite
$\lambda$ solutions as a function of density, and the associated
errors bars are indicated.
The hatched region indicates a transition region
from finite to zero $\lambda$, and it width is $2\sigma$, where
sigma is the error assigned to the transition density estimate.}}

\vbox to 3.5in{\vss\hbox to 5.625in{\hss
{\includegraphics{8qfig2.ps}}\hss}}
\nobreak
{\noindent\narrower{{\bf FIG.~\protect{\ref{fig8qfig2}}}.
The energy per particle
$\varepsilon(\lambda)$ as a function of $\lambda$ for
various, fixed $\rho$. The lines are merely to guide the eye. The
solid lines are $\varepsilon(\lambda)$ versus $\lambda$ in the
transition region to zero $\lambda$; the densities for those curves
from bottom to top are .90, .95, and 1.0, respectively.}}
\bigskip

  Figure {\ref{fig8qfig3}} shows the energy per particle of the eight
quark system for the optimum variational parameters of Fig.~{\ref{fig8qfig1}}.
The energies of the finite $\lambda$ and zero $\lambda$ solutions
are plotted in the same figure. As previously illustrated in
Fig.~{\ref{fig8qfig2}}, the energies of the finite and zero
$\lambda$ solutions cross at $\rho\sim 1$. As the density increases, the
potential energy goes to zero. In that limit, the total energy per
particle is that of a free Fermi gas.
This is natural -- the
interquark separation,
and, hence, the string length,
approaches zero as $\rho\rightarrow\infty$. At low densities, the
total energy per particle approaches the isolated hadron value, and
the virial theorem ($t=v$) appears to be satisfied in that limit as well.
Note that nuclear matter in this eight quark calculation, at least,
is not bound.
At moderate densities, a dynamical
interplay exists between the potential and
kinetic energies. That is, for a slightly larger average string length,
the shape of the
spatial wavefunction can ``relax'' and reduce its kinetic energy.

We shall now
consider the variational parameter $\lambda$ and the energy per
particle as a function of density for twelve quarks, as we would like
to see how sensitive our results are to finite size effects.
Figure {\ref{fig12qfig1}} shows the twelve quark results for
the optimal variational parameter $\lambda$ as a function of $\rho$.
The eight quark results from Fig.~{\ref{fig8qfig1}} are plotted as well
for comparison. The twelve quark calculation approaches the
isolated hadron limit as $\rho\rightarrow 0$, and it undergoes a
transition to zero $\lambda$ in the same region of $\rho$. However,
the results at intermediate density are different. The optimum
variational parameter in the twelve quark case decreases more quickly as the
system moves away from the isolated hadron limit,
although it does not have the dip at $\rho\sim .75$ that exists in
the eight quark case.
If we now
turn to the energy per particle as a function of density, we see
the finite size effects are substantial there as well. These results
are shown in Fig.~{\ref{fig12qfig2}}. The energy per particle for the

\vbox to 3.75in{\vss\hbox to 5.625in{\hss
{\includegraphics{8qfig3.ps}}\hss}}
\nobreak
{\noindent\narrower{{\bf FIG.~\protect{\ref{fig8qfig3}}}.
The total energy per particle
$\varepsilon(\rho)$ as a function of $\rho$ for eight quarks.
The kinetic energy per particle is denoted by ``$t$'', whereas
the potential energy per particle is denoted by ``$v$''. The lines
are merely to guide the eye. The points connected by lines are
the energies associated with the finite $\lambda$ variational solution.
The energies $\varepsilon$, $t$, and $v$ associated with the zero $\lambda$
solution are plotted as well; the symbol definitions are indicated in
the figure. The error bars associated with these points are plotted,
when they are large enough to be visible. The $t(\lambda=0)$ points
correspond to a free Fermi gas, so that there are no associated errors.
}}
\bigskip

\noindent{twelve quark case obeys the high and low density limits of}
the model: the system
evolves to the isolated hadron energy per
particle and to equal $t$ and $v$ as the density goes to zero, and it
approaches the Fermi gas limit as the density goes to infinity.
Figure {\ref{fig12qfig1}} also shows the level crossing behavior of the
finite and zero $\lambda$ families seen in Fig.~{\ref{fig8qfig1}} and inferred
from Fig.~{\ref{fig12qfig1}}. The ``softening'' of
$\lambda$ seen in the twelve quark case at $\rho\sim .3$ is reflected
in the slower increase in $t$ in that density regime. The consequence
is that the energy per particle skims along at the isolated hadron
energy per particle for a non-trivial interval in $\rho$ before finally
increasing. This is in contrast to the eight quark case, where $\varepsilon$
increases steadily as a function of density.
Neither the eight nor
twelve quark case supports a bound state; yet,
the change in $\varepsilon$
in going from eight to twelve quarks in the $\rho\sim .3$ regime suggests
that a bound state can not be ruled out in this model as the
number of quarks goes to infinity.
The finite size effects are still
larger in the $\rho\sim .75$ regime. There, the interplay between $t$
and $v$ seen in the eight quark case is simply absent. This is
entirely consistent with the behavior of $\lambda$ in this density
regime. The twelve quark $\lambda$ decreases smoothly
to the transition region -- the dip behavior seen in the eight quark
case is absent.

\vbox to 3.0in{\vss\hbox to 5.625in{\hss
{\includegraphics{12qfig1.ps}}\hss}}
\nobreak
{\noindent\narrower{{\bf FIG.~\protect{\ref{fig12qfig1}}}.
The variational parameter $\lambda$ as a function
of $\rho$ for twelve quarks. The eight quark results from
Fig.~\protect{\ref{fig8qfig1}} are shown as well, so that the
the finite size effects may be seen more clearly. The twelve
quark data is denoted by ``$\times$''.
The points indicate the optimum finite
$\lambda$ solutions as a function of density, and the associated
errors bars are indicated.
The hatched region indicates a transition region
from finite to zero $\lambda$, and it width is $2\sigma$, where
sigma is the error assigned to the transition density estimate.
This estimate is identical to that of the eight quark case.}}
\bigskip
\bigskip
\nobreak
\vbox to 3.5in{\vss\hbox to 5.625in{\hss
{\includegraphics{12qfig2.ps}}\hss}}
\nobreak
{\noindent\narrower{{\bf FIG.~\protect{\ref{fig12qfig2}}}.
The energy per particle $\varepsilon(\rho)$ as a function of $\rho$ for
twelve quarks. The notation is as per Fig.~\protect{\ref{fig8qfig3}}.}
\vfill
\eject

\vbox to 3.5in{\vss\hbox to 5.625in{\hss
{\includegraphics{figpaint1.ps}}\hss}}
\nobreak
{\noindent\narrower{{\bf FIG.~\protect{\ref{figpaint1}}}.
The optimum variational parameter $\lambda$ in the painted
model, Eq.~(\protect{\ref{paintans}}), as a function of
density for eight and twelve quarks. The curves are merely
to guide the eye. The dashed curve connects the eight quark
calculations, whereas the solid curve connects the twelve
quark ones.
}}
\bigskip
\vbox to 3.5in{\vss\hbox to 5.625in{\hss
{\includegraphics{figpaint2.ps}}\hss}}
\nobreak
{\noindent\narrower{{\bf FIG.~\protect{\ref{figpaint2}}}.
The energy per particle as a function of density in the painted model for
eight and twelve quarks. The curves are merely to guide
the eye. As in Fig.~\protect{\ref{figpaint1}}, the dashed
line connects the eight quark calculations, and the solid
line connects the twelve quark ones. The statistical errors
that are not shown are negligible relative to the scale of
the figure. The remaining notation is as in
Fig.~\protect{\ref{fig8qfig3}}.
}}
\bigskip
\vfill
\eject

We can now compare the results of the full $SU(2)$ calculation,
which we have just presented, to the results of the approximate
$SU(2)_c$ treatment -- the ``painted model'' described in
Sec.~{\ref{conferre}}. Figure {\ref{figpaint1}} shows the
optimum variational parameter in the painted model
as a function of density for eight
and twelve quarks. Here, as above, the variational parameter
approaches the isolated hadron limit at low densities. At high
densities, however, $\lambda$ flows smoothly to zero; there is
no abrupt transition to zero $\lambda$ in this model.
The finite size effects are small for $\rho\lapp.6$; they only
become significant at higher densities.
If finite size effects
are to exist in an isolated density regime, then it is sensible that
they should exist only at larger densities.
The variational
parameters of the models do {\it not} bear closer comparison; the
{\it Ans\"atze} from whence they are derived are decidedly different.

The energy per particle in the painted model is shown as a function
of density in Fig.{\ref{figpaint2}} for eight and twelve quarks.
The model results approach the isolated hadron limit at low
density and the free Fermi gas limit at high density. As above,
the potential energy tends slowly to zero as the density increases.
Here the finite size effects are small -- even at larger
densities. The total energies per particle in the
magic and painted models are compared in Fig.{\ref{figenergy}}.
The twelve quark calculations in the two models are fairly
similar, even
though the physics of the models is rather different.
In comparing Figs.{\ref{fig12qfig2}} and {\ref{figpaint2}}, we note that
differences in $t$ and $v$ exist, but

\vbox to 3.5in{\vss\hbox to 5.625in{\hss
{\includegraphics{figenergy.ps}}\hss}}
\nobreak
{\noindent\narrower{{\bf FIG.~\protect{\ref{figenergy}}}.
The energy per particle in the
magic (full $SU(2)$) and painted (approximate $SU(2)$) models
as a function of density for
eight and twelve quarks. The starred data
points joined by the solid curve are
the twelve quark painted model calculations from
Fig.~\protect{\ref{figpaint2}}. The other solid curve connects
the twelve quark magic model calculations from
Fig.~\protect{\ref{fig12qfig2}}. The dashed curve joins the corresponding
eight quark calculations from Fig.~\protect{\ref{fig8qfig3}}.
Note that the energy per particle in the full $SU(2)$ calculation is
for the optimal variational parameter solution -- $\lambda$ is not
required to be finite.
}}
\bigskip

\noindent{they tend to compensate. The}
similarity of the total energy per particle in the twelve
quark cases does tend to support the use of the painted model
for calculations of this ilk.
It would be interesting to see if this similarity survives in
calculations with more quarks.

\subsection{GROUND STATE STRUCTURE}

  We shall now turn to a study of the system's ground state structure
as a function of density. Figures {\ref{fig8qcomp}} and
{\ref{fig12qcomp}} illustrate the two-body density versus the interquark
separation $r$ for various one-body densities in the eight and
twelve quark cases. Both the calculations in the full and
approximate $SU(2)_c$ models are shown; we hope that their
comparison will yield further insight into the structure of the
ground state.

\vbox to 4.75in{\vss\hbox to 5.625in{\hss
{\includegraphics{fig8qcomp.ps}}\hss}}
\nobreak
{\noindent\narrower{{\bf FIG.~\protect{\ref{fig8qcomp}}}.
The two-body density as a function of the interquark separation
$r$ for a variety of densities for the eight quark case. Both
the results of the full (magic) and approximate $SU(2)$ (painted) calculations
have been plotted. The full $SU(2)$ points are joined by a solid
line to guide the eye. The dashed curve is the two-body density
appropriate to a free Fermi gas (Eq.~(\protect{\ref{rho2high}})),
and the dotted curve shows
that appropriate to the isolated hadron limit
(Eq.~(\protect{\ref{rho2low}})).
Note that
in (a) $\rho=.25$, in (b) $\rho=.50$, in (c) $\rho=.95$ and
$\lambda$ is the optimum finite solution, and in (d) $\rho=.95$
and $\lambda=0$.
}}
\bigskip

\vbox to 4.0in{\vss\hbox to 5.625in{\hss
{\includegraphics{fig12qcomp.ps}}\hss}}
\nobreak
{\noindent\narrower{{\bf FIG.~\protect{\ref{fig12qcomp}}}.
The two-body density as a function of the interquark separation
$r$ for a variety of densities for the twelve quark case.
The notation is identical to that of Fig.~\protect{\ref{fig8qcomp}}.
}}
\bigskip

\noindent{We shall consider the eight quark results,}
Fig.~{\ref{fig8qcomp}}, first. Figure {\ref{fig8qcomp}} shows the
two-body density from low to high ((a)$\rightarrow$(d)) density. Both the
magic and painted models approach the isolated hadron limit at
low density, as they should, although the painted model approaches
the limit more quickly as $\rho$ decreases. Note that in all cases
$\rho_2(r)$ approaches one for sufficiently large $r$.
The two-body density for
the full $SU(2)_c$ case is shown at still lower density in
Fig.~{\ref{figlowrho}(a)}; $\rho_2(r)$ does approach the isolated
hadron limit at low density, as claimed.
The painted model's more
rapid approach to the isolated hadron limit can be readily understood
through analyzing the models' {\it Ans\"atze}. Equations
(\ref{x13magic}) and (\ref{x13paint}) show the structure of the two
ground states as two quarks of the same color approach each other.
The healing of the Fermi wound in the magic model contains
an additional $L$-independent term; this is likely why the painted model
approaches the isolated hadron limit more quickly.
At moderate density, $\rho=.5$, the full and approximate $SU(2)$
calculations are quite different, although
there is no reason why they should be the same.
At still higher densities, the
two models are again similar, but that
is because they are approaching the Fermi gas limit. At $\rho=.95$,
$\lambda$ in the painted model is small, but finite, and the
corresponding two-body density is nearly that of a free Fermi gas.
The same is true of the magic model, although $\lambda$ exhibits a
rapid drop to zero in this density regime. The finite $\lambda$
two-body density in (c) is only slightly
different from
the  zero $\lambda$ result in (d); apparently the structure of the
state changes little across the ``transition.'' The result in (d)
is essentially identical to the free Fermi gas result; this implies
that the finite-size effects are under some measure of
control. We can explore this issue directly by studying the twelve
quark results in Fig.~{\ref{fig12qcomp}}.

The finite-size effects in $\rho_2(r)$ under
the change from eight to twelve quarks are significant in the magic
model, although they are modest in the painted model. Note that
the zero $\lambda$ two-body density in (d) shows reasonable agreement
with the free Fermi gas result: the finite-size effects are in the
finite $\lambda$ calculation. Perhaps the {\it Ansatz} of
Eq.~(\ref{8qansatz}) is ineffective in describing the additional correlation
the quarks feel due to confinement at moderate density. In particular,
the two-body density now changes as $\lambda$ changes
from its optimum finite value to zero at the transition density.
In general, $\rho_2(r)$ in the magic model has additional strength at
$r=0$, compared to the eight quark results. This calculation
still appears to satisfy the isolated hadron limit, however, as shown in
Fig.~{\ref{figlowrho}(b)}. The ground state structure of the painted
and magic models at the twelve quark level appears to be rather different.
As in the eight quark case, the painted model approaches the isolated
hadron limit much more rapidly. The differing structure of the
two models' {\it Ans\"atze},
Eqs.~(\ref{x13magic}) and (\ref{x13paint}), as

\vbox to 3.0in{\vss\hbox to 5.625in{\hss
{\includegraphics{figlowrho.ps}}\hss}}
\nobreak
{\noindent\narrower{{\bf FIG.~\protect{\ref{figlowrho}}}.
The two-body density as a function of the interquark separation
$r$ for the full $SU(2)_c$ calculation -- the ``magic'' model --
in the low density limit. Case (a) is for eight quarks with
$\rho=.08$, and case (b) is for twelve quarks with $\rho=.1$.
In each case, the dotted line corresponds to the
appropriate isolated hadron two-body density.
}}
\bigskip

\noindent{two quarks of the same color approach each other,}
is likely responsible for this effect. The nature of the correlations
included in the painted and magic model are different, and the two-body
density is apparently quite sensitive to those differences. It would
be interesting to see whether this sensitivity would be preserved
in calculations performed with still more quarks.

The approach of the magic model to the isolated hadron limit, as
shown in Fig.~{\ref{figlowrho}}, serves to illustrate another
point concerning the structure of the state. At small $r$, the Monte
Carlo calculations of the two-body density track the isolated hadron
results fairly closely. At larger $r$, however, there is a significant
``filling-in'' of the two-body density relative to the isolated
hadron result. This is in contrast to the painted model calculations at
low density, as depicted in Figs.~{\ref{fig8qcomp}}(a) and
{\ref{fig12qcomp}}(a). There the two-body density tracks the isolated
hadron result until it reaches essentially zero; it eventually
picks up again because other hadron clusters exist at finite density.
The difference in the behavior of the two models as $r$ increases
in the low density limit arises from the exchange terms included
in the magic model. For any fixed, finite density, there exists
some $r$  for which the exchange term to recouple a quark pair
into different hadrons is finite. Such terms do not exist in
the painted model as the colors of the quarks are fixed.
This effect
explains the ``wings'' seens on the two-body density in the magic
model shown in Fig.~{\ref{figlowrho}}.

We can understand the difference between the magic and painted
models at low densities in yet another way.
That is, the large dip below one
in the painted model's $\rho_2(r)$  at large
$r$
can be interpreted in terms of a repulsive effective force between hadron
clusters in this model. This is simply due to antisymmetry. A red
quark in one cluster must be anticorrelated with the red quark it sees
in an adjacent cluster. In the magic model, the inclusion of
exchange terms allows some additional attraction to exist, since the
colors of the quarks can change. As a red quark in one cluster
approaches a red quark in an adjacent cluster, the first quark can flip
its color and pair off with the red quark it approaches. In this
way the effective attraction -- relative to the painted model --
is generated.

We have now completed our survey of results in the full
$SU(2)_c$ model.



\section{SUMMARY AND OUTLOOK}
\label{sumup}

  Here we have considered nuclear matter in the context of a
non-relativistic constituent quark model with an internal
color degree of freedom. In part, we have
been motivated by a desire to understand traditional issues,
such as nuclear matter binding and saturation, from a quark
model viewpoint. Our primary interest, however, is in the properties of
the system as a function of density. The model system we consider
becomes a gas of isolated hadrons at low density, and a
non-relativistic free Fermi gas at high density. Thus, it provides
a natural laboratory in which to understand how the combined effects
of confinement and antisymmetry determine the structure of the
system as it evolves between the two limits. In particular, we
wish to gain insight into a possible ``deconfinement'' phase
transition at high density.

 The model we consider is a simple one. It is derived from the
combined application of the adiabatic and strong coupling
limits to QCD, and it consists of the following prescription:
the quarks are always paired to the lowest energy set of
flux tubes. Consequently, the model contains no dynamical
gluon degrees of freedom; the
dynamics are those of quark exchange. We remind the reader that
the global minimization prescription which defines our model
also guarantees that it is free from residual color
Van der Waals forces -- an advantage over
two-body potential models \cite{lipkin}.
The model is non-relativistic,
although we are, in principle, interested in density regimes in
which relativistic effects could be important. We feel that
the model's simplicity and intrinsic appeal
outweighs its disadvantages for the
qualitative purposes of our study. Its primary advantage is
that it places quark and hadron degrees of freedom on an equal
footing. The structure of the system at intermediate densities
is due entirely to the model dynamics, namely of
confinement and antisymmetry, rather than
to the heuristic approximations obliged in a more complicated model.

  The quark exchange model has been used in previous studies
of nuclear matter \cite{hmn,cjhjp1d,cjhjp3d,watson,wanda,frichter}.
Our contribution here has been the complete incorporation of an internal
$SU(2)$ color degree of freedom, although we have continued
to ignore the quarks' spin. By ``complete'', we mean
that we have incorporated the internal degree of freedom without
giving the quarks {\it fixed} colors. This is important: QCD requires
that a many-quark system be an overall color singlet, but it does not
determine the colors of the individual quarks.
We have constructed an explicitly antisymmetric variational
{\it Ansatz} with the internal degree
of freedom, and then proceeded with a Monte Carlo calculation to
determine the properties of the system.

  We have found that the
$SU(2)_c$ results look rather different from earlier ``spin 0 fermion''
studies \cite{hmn,cjhjp1d}.
Indeed, the simulations support an abrupt transition from
finite to zero variational parameter, which describes
the system's clustering into hadrons, at moderately high density.
The system with zero variational parameter is a free Fermi gas.
The change of the structure of the system across this transition
is sensitive to finite-size effects. At the eight quark level,
the two-body density changes little as the variational parameter
changes from its optimum finite value to zero at the transition
density. However, the same calculation performed at the twelve
quark level shows a significant effect. A definitive conclusion
as to the structure of the system as it goes through the transition
regime must be reserved, then, for later study. This abrupt change
in the variational pararmeter as a function of density is also
absent in the ``painted'' -- the quarks have fixed colors --
$SU(2)$ study we considered here for comparison. It would be
interesting to see if this effect persists in three spatial
dimensions and in full $SU(3)$ color.

  The energy per particle as a function of density in this model shows rich
behavior as a function of density.
At moderate densities, the system is capable of
tolerating longer string lengths in order to reduce its kinetic energy.
We should note, in passing, that in a model with linear, rather than
harmonic, confinement these effects would likely be more pronounced.
  The energy per particle rises more slowly with density
than those of earlier studies without color, but nuclear matter
in this model is not bound. The energy per particle
does become considerably flatter with density under the change from
eight to twelve quarks, however. The change is, in fact, big enough that
the existence of a bound state can not
be ruled out as the number of quarks increases.
Yet, the lack of a bound state is hardly a surprise.
The dynamics in this model are
simple and do not suffice to reproduce the known features -- such as the
spin dependence -- of the $N-N$ force.

  Unlike the full $SU(2)$ model, the
painted $SU(2)$  calculation of the energy per particle shows
small finite size effects under the change from eight to twelve quarks.
The twelve quark energy per particle
results in the full and painted $SU(2)$ calculations
are surprisingly similar. This is in contrast to the evolution of the
variational parameter and two-body density of the two calculations
with density, which are different. It would be interesting to see
whether this similarity persists in a calculation with more quarks.
The similarity which does exist, however, lends support to the further
use of the painted model in exploring the energetics of bulk quark
matter, for example, when flavor degrees of
freedom are included \cite{jorgep}. However, the distinctly
different two-body densities in the two models indicate that
a complete treatment of antisymmetry is necessary to understanding
the system's ground state structure. Indeed, this may also
indicate that a complete incorporation of spin and flavor
degrees of freedom as well are essential to an understanding of the
ground state of hadronic matter with density.

 The computation of the two-body density in this model has given
us insight into the structure of the ground state as a function of
density. In principle, we would also like to study the excitations
of the system as well. We can explore the ``phonon-like'' behavior
of the system through the calculation of the Fourier transform of the
two-body density \cite{feynman}.
It is possible, however, to gain access to the full
response of the system. That is, we can calculate the quark response
function in imaginary time, and then reconstruct the longitudinal
response function, familiar from  inclusive electron scattering, by
finding the inverse Laplace transform of the Euclidean response.
The inversion procedure is non-unique, but
Carlson and Schiavilla \cite{carlson}
have shown it practicable for light nuclei.
We imagine that the quark response function would evince rich behavior
as a function of density.


\acknowledgements

We thank B. Serot for a critical reading of the manuscript.
This research was supported
by the AAUW Educational Foundation (S.~G.), by the
DOE under contracts \# DE-AC05-84ER40150 (S.~G.) and
\# DE-FG02-87ER40365 (S.~G. and C.~J.~H.), and by the
Florida State University Supercomputer Computations Research Institute
through the DOE contracts \#
DE-FC05-85ER250000 and DE-FG05-92ER40750 (J.~P.).


\begin{figure}
\caption{
An illustration of quark exchange dynamics for 4 quarks.
Two incoming hadrons with quark content $(12)_0$
and $(34)_0$ may exchange quarks -- the
strings connecting 1 to 2  and 3 to 4 flip --
to yield a $(13)_0(24)_0$
outgoing state. }
\label{figtahiti}
\end{figure}
\begin{figure}
\caption{
The variational parameter $\lambda$ as a function
of $\rho$ for eight quarks. The points indicate the optimum finite
$\lambda$ solutions as a function of density, and the associated
errors bars are indicated.
The hatched region indicates a transition region
from finite to zero $\lambda$, and it width is $2\sigma$, where
sigma is the error assigned to the transition density estimate.}
\label{fig8qfig1}
\end{figure}
\begin{figure}
\caption{
The energy per particle
$\varepsilon(\lambda)$ as a function of $\lambda$ for
various, fixed $\rho$. The lines are merely to guide the eye. The
solid lines are $\varepsilon(\lambda)$ versus $\lambda$ in the
transition region to zero $\lambda$; the densities for those curves
from bottom to top are .90, .95, and 1.0, respectively.}
\label{fig8qfig2}
\end{figure}
\begin{figure}
\caption{
The total energy per particle
$\varepsilon(\rho)$ as a function of $\rho$ for eight quarks.
The kinetic energy per particle is denoted by ``$t$'', whereas
the potential energy per particle is denoted by ``$v$''. The lines
are merely to guide the eye. The points connected by lines are
the energies associated with the finite $\lambda$ variational solution.
The energies $\varepsilon$, $t$, and $v$ associated with the zero $\lambda$
solution are plotted as well; the symbol definitions are indicated in
the figure. The error bars associated with these points are plotted,
when they are large enough to be visible. The $t(\lambda=0)$ points
correspond to a free Fermi gas, so that there are no associated errors. }
\label{fig8qfig3}
\end{figure}
\begin{figure}
\caption{
The variational parameter $\lambda$ as a function
of $\rho$ for twelve quarks. The eight quark results from
Fig.~\protect{\ref{fig8qfig1}} are shown as well, so that the
the finite size effects may be seen more clearly. The twelve
quark data is denoted by ``$\times$''.
The points indicate the optimum finite
$\lambda$ solutions as a function of density, and the associated
errors bars are indicated.
The hatched region indicates a transition region
from finite to zero $\lambda$, and it width is $2\sigma$, where
sigma is the error assigned to the transition density estimate.
This estimate is identical to that of the eight quark case.}
\label{fig12qfig1}
\end{figure}
\begin{figure}
\caption{
The energy per particle $\varepsilon(\rho)$ as a function of $\rho$ for
twelve quarks. The notation is as per Fig.~\protect{\ref{fig8qfig3}}.}
\label{fig12qfig2}
\end{figure}
\begin{figure}
\caption{
The optimum variational parameter $\lambda$ in the painted
model, Eq.~(\protect{\ref{paintans}}), as a function of
density for eight and twelve quarks. The curves are merely
to guide the eye. The dashed curve connects the eight quark
calculations, whereas the solid curve connects the twelve
quark ones.
}
\label{figpaint1}
\end{figure}
\begin{figure}
\caption{
The energy per particle as a function of density in the painted model for
eight and twelve quarks. The curves are merely to guide
the eye. As in Fig.~\protect{\ref{figpaint1}}, the dashed
line connects the eight quark calculations, and the solid
line connects the twelve quark ones. The statistical errors
that are not shown are negligible relative to the scale of
the figure. The remaining notation is as in
Fig.~\protect{\ref{fig8qfig3}}.
}
\label{figpaint2}
\end{figure}
\begin{figure}
\caption{
The energy per particle in the
magic (full $SU(2)$) and painted (approximate $SU(2)$) models
as a function of density for
eight and twelve quarks. The starred data
points joined by the solid curve are
the twelve quark painted model calculations from
Fig.~\protect{\ref{figpaint2}}. The other solid curve connects
the twelve quark magic model calculations from
Fig.~\protect{\ref{fig12qfig2}}. The dashed curve joins the corresponding
eight quark calculations from Fig.~\protect{\ref{fig8qfig3}}
Note that the energy per particle in the full $SU(2)$ calculation is
for the optimal variational parameter solution -- $\lambda$ is not
required to be finite.
}
\label{figenergy}
\end{figure}
\begin{figure}
\caption{
The two-body density as a function of the interquark separation
$r$ for a variety of densities for the eight quark case. Both
the results of the full (magic) and approximate $SU(2)$ (painted)
calculations
have been plotted. The full $SU(2)$ points are joined by a solid
line to guide the eye. The dashed curve is the two-body density
appropriate to a free Fermi gas (Eq.~(\protect{\ref{rho2high}})),
and the dotted curve shows
that appropriate to the isolated hadron limit
(Eq.~(\protect{\ref{rho2low}})).
Note that
in (a) $\rho=.25$, in (b) $\rho=.50$, in (c) $\rho=.95$ and
$\lambda$ is the optimum finite solution, and in (d) $\rho=.95$
and $\lambda=0$.
}
\label{fig8qcomp}
\end{figure}
\begin{figure}
\caption{
The two-body density as a function of the interquark separation
$r$ for a variety of densities for the twelve quark case.
The notation is identical to that of Fig.~\protect{\ref{fig8qcomp}}.
}
\label{fig12qcomp}
\end{figure}
\begin{figure}
\caption{
The two-body density as a function of the interquark separation
$r$ for the full $SU(2)_c$ calculation -- the ``magic'' model --
in the low density limit. Case (a) is for eight quarks with
$\rho=.08$, and case (b) is for twelve quarks with $\rho=.1$.
In each case, the dotted line corresponds to the
appropriate isolated hadron two-body density.
}
\label{figlowrho}
\end{figure}
%



\begin{references}
%
\bibitem[\dag]{svg}e-mail:  gardner@iucf.indiana.edu
%
\bibitem[\ddag]{cjh}e-mail: charlie@iucf.indiana.edu
%
\bibitem[\S]{jp}e-mail: jorgep@scri.fsu.edu
%
\bibitem{matsuisatz} T.\ Matsui and H.\ Satz,
Phys.\ Lett.\ {\bf 178B}, 416 (1986).
%
\bibitem{gavin} S.\ Gavin, M.\ Gyulassy, A.\ Jackson,
Phys.\ Lett.\ {\bf 207B}, 257 (1988).
%
\bibitem{kurihara} J.\ H\"ufner, Y.\ Kurihara, H.\ J.\ Pirner,
 Phys.\ Lett.\ {\bf 215B}, 218 (1988).
%
\bibitem{lenz} F.\ Lenz, J.\ T.\ Londergan, E.\ J.\ Moniz,
R.\ Rosenfelder, M.\ Stingl, and K.\ Yazaki, Ann.\ Phys.\
{\bf 170}, 65 (1986).
%
\bibitem{hmn} C.\ J.\ Horowitz, E.\ J.\ Moniz, and J.\ W.\ Negele,
 Phys.\ Rev.\ {\bf D31}, 1689 (1985).
%
\bibitem{cjhjp1d} C.\ J.\ Horowitz and J.\ Piekarewicz,
 Phys.\ Rev.\ {\bf C44}, 2753 (1991).
%
\bibitem{cjhjp3d} C.\ J.\ Horowitz and J.\ Piekarewicz,
 Nucl.\ Phys.\ {\bf A536}, 669 (1992).
%
\bibitem{watson} P.\ J.\ S.\ Watson, Nucl.\ Phys.\ {\bf A494}, 543 (1989);
A.\ B.\ Migdal and P.\ J.\ S.\ Watson, Phys.\ Lett.\ {\bf 252B}, 32 (1990).
%
\bibitem{wanda} W.\ M.\ Alberico, M.\ B.\ Barbaro, A.\ Molinari,
and F.\ Palumbo, Z.\ Phys.\ {\bf A341}, 327 (1992);
W.\ M.\ Alberico, M.\ B.\ Barbaro, A.\ Magni, and M.\ Nardi,
 Nucl.\ Phys.\ {\bf A552}, 495 (1993).
%
\bibitem{frichter} G.\ M.\ Frichter and J.\ Piekarewicz,
preprint \# FSU-SCRI-93-61,
accepted for publication in Computers in Physics.
%
\bibitem{gardner} S.\ Gardner and E.\ J.\ Moniz, Phys.\ Rev.\
{\bf C36}, 2504 (1987); S.\ Gardner, Phys.\ Rev.\ {\bf C42}, 2193 (1990).
%
\bibitem{lissia} M.\ Lissia and J.\ W.\ Negele, Phys.\ Rev.\
{\bf D39}, 1413 (1989).
%
\bibitem{feinberg} G.\ Feinberg and J.\ Sucher, Phys.\ Rev.\
{\bf D20}, 1717 (1979).
%
\bibitem{lipkin} O.\ W.\ Greenberg and H.\ J.\ Lipkin,
Nucl.\ Phys.\ {\bf A370}, 349 (1981).
%
\bibitem{paton} N.\ Isgur and J.\ Paton, Phys.\ Rev.\ Lett.\ {\bf 54},
869 (1985).
%
\bibitem{isgur} N.\ Isgur, in {\it The New Aspects of Subnuclear Physics},
edited by A.\ Zichichi (Plenum, New York, 1980).
%
\bibitem{harvey} M.\ Harvey, J.\ Letourneux, and B.\ Lorazo,
Nucl.\ Phys.\ {\bf A424}, 428 (1984).
%
\bibitem{maltman} K.\ Maltman and N.\ Isgur,
Phys.\ Rev.\ {\bf D29}, 952 (1984).
%
\bibitem{econo} R.\ E.\ Burkard and U.\ Derigs, {\it Lecture
Notes in Economics and Mathematical Systems} (Springer-Verlag,
Berlin, 1980), vol. 184.
%
\bibitem{isgurkarl} N.\ Isgur and G.\ Karl, Phys.\ Rev.\ {\bf D20},
1191 (1979).
%
\bibitem{jorgep} J.\ Piekarewicz, in preparation.
%
\bibitem{feynman} R.\ P.\ Feynman, {\it Statistical Mechanics},
(Addison-Wesley, Reading, 1972),
p.328{\it ff}.
%
\bibitem{carlson} J.\ Carlson and R.\ Schiavilla, Phys.\ Rev.\ Lett.\
{\bf 68}, 3682 (1992).
%


\end{references}
\end{document}